\documentclass[12pt]{article} % 10pt is ignored!

\usepackage{epsfig}
\usepackage{graphicx}% Include figure files
\usepackage{amsmath}
\usepackage{amssymb}
\usepackage{url}

\begin{document}

\title{Modal series expansions for plane gravitational waves}

% The list of authors, and the short list which is used in the headers.
% If you need two or more lines of authors, add an extra line using \newauthor
\author{L. Acedo\thanks{E-mail: luiacrod@imm.upv.es}\\
Instituto Universitario de Matem\'atica Multidisciplinar,\\
Building 8G, $2^{\mathrm{o}}$ Floor, Camino de Vera,\\
Universitat Polit$\grave{\mbox{e}}$cnica de Val$\grave{\mbox{e}}$ncia,\\
Valencia, Spain\\
}

\maketitle

\begin{abstract}

Propagation of gravitational disturbances at the speed of light is one of the key predictions of the General Theory of Relativity. This result is now backed indirectly by the observations of the behaviour of the ephemeris of binary pulsar systems. 
These new results have increased the interest in the mathematical theory of gravitational waves in the last decades and 
several mathematical approaches have been developed for a better understanding of the solutions.

In this paper we develop a modal series expansion technique in which solutions can be built for plane waves from a
seed integrable function. The convergence of these series is proven by the Raabe-Duhamel criteria and we show that these solutions are characterized by a well-defined and finite curvature tensor and also a finite energy content.

\end{abstract}


\maketitle

\section{Introduction}
\label{sec_1}

The mathematical theory of gravitational waves has a troubled history. Confusion among curvature and coordinate
singularities lead Einstein and Rosen to publish a paper in which they claimed that gravitational radiation 
does not exist \cite{EinsteinRosen}. This question was still pondered by Rosen as late as 1979 \cite{Rosen1979}. 
Nowadays, the indirect evidence obtained by the meticulous tracking of the stars in the PSR B1913$+$16 system has provided bedrock evidence of the exactitude of the quadrupole formula for the gravitational energy losed by a binary system in the form of gravitational waves \cite{Sivaram,Weisberg2010}. 

The confidence contributed by this experimental evidence has encouraged a lot of research into the mathematical properties
of isolated and colliding gravitational waves in the last decades \cite{Griffiths}. In the case of plane gravitational waves
the differential equation corresponds to the solution of the Sturm-Liouville problem:

\begin{equation}
\label{SL}
p''(u)+V(u)\, p(u)=0\; ,
\end{equation} 

where the apostrophes denote the second-order derivative, $p(u)$ is a metric function and $V(u)$ is the fourth Weyl scalar, $\Psi_4$ \cite{Chandrasekhar}. Exact solutions in terms of special functions can be obtained for special selections of the fuction $V(u)$. However, this is not possible in general and one must resort to numerical integration or Taylor series in powers
of $u$. Although approximate solutions with a certain domain of validitiy can be obtained with these traditional techniques, the main hurdle is the difficulty to prove convergence and finite energy content with numerical calculations. In this context, Bini
et al. \cite{Bini} proposed some constraints that a realistic gravitational wave should verify: (i) The Riemann tensor exists everywhere and it is continuous (ii) Space-time is asymptotically flat for $u=x-ct \rightarrow \pm \infty$ (iii) The metric function is regular (iv) The gravitational wave carries a finite energy content. We will show that the method described in this paper is
adequate to obtain a family of solutions verifying these properties. 

Recently, it has been found that certain classes of non-linear differential equations with quadratic
terms can be solved analytically by means of a modal series expansion. This method has been successfully applied to a version
of the Kermack-McKendrick model with mass action terms \cite{AcedoNARWA,AcedoPhysA} and also to the Lorenz system in the laminar
flow regime \cite{Acedo2013}. This method can be described in general as follows: for any non-linear operator $\mathcal{R}$ acting upon a function $g$ we may define a nonlinear differential equation:

\begin{equation}
\label{difeg}
\mathcal{R}(g)=0\; .
\end{equation}

A tentative series solution for this equation is proposed in terms of the functional series $\mathcal{F}=\left\{ f_n \right\}$, $n=0,1,2,\ldots$ in the form:

\begin{equation}
\label{genser}
g = \sum_{n=0}^\infty \,  a_n f_n \; ,
\end{equation}

where $a_n$ are real or complex numbers. The series in Eq.\  (\ref{genser}) is called modal under the nonlinear operator $\mathcal{R}$ if  $\mathcal{R}(f_m)$ can be written as an algebraic combination of a finite subset of functions in $\mathcal{F}$ 
for any integer $m$. This allows us to find a recurrence relation for the coefficients  $a_n$, $n=0,1,2,\ldots$. The number of free
coefficients in Eq.\ (\ref{genser}) will correspond to the highest-order derivative in the operator $\mathcal{R}(f_m)$. In the case
of General Relativity the equations are order two in the metric. Sometimes the recurrence relation involve terms with indexes larger than the one we are computing and the calculation cannot be closed. Even if an adequate recurrence is found, we must test
the final series solution for convergence because this is not assured from the outline of the heuristic modal method.

We will develop the modal series technique for plane gravitational waves to show that exact solutions with finite values of the metric and the curvature exist. These waves form an infinite plane front perpendicular to the the propagation axis, $x$, and the Minkowski's metric is recovered in the asymptotic limits $x\rightarrow \pm \infty$.

The structure of the paper is as follows: in Section \ref{sect_2} we develop the modal formalism for gravitational plane waves.  In Section \ref{sect_3} we prove the 
convergence of these solutions for any coordinates and discuss the definition of the curvature. The stress-energy pseudotensor
is explicitly calculated in Section \ref{sect_4}. We also prove here that the energy content is finite.
 Conclusions, remarks and prospective future work are discussed in Section \ref{sect_5}.

\section{Modal transseries for gravitational waves}
\label{sect_2}

We consider the metric for a plane-wave in the Rosen's form:
\begin{equation}
\label{metric}
d s^2=-c^2 dt^2+d x^2+p^2(u)\, d y^2+q^2(u)\, d z^2= 2\, d u\, d v + p^2(u) \, d y^2 + q^2(u) \, d z^2 \; ,
\end{equation}
where $u=x-c t$ and $v=x+c t$. Einstein's field equations in vacuum for this metric \cite{Rindler} are deceptively simple:
\begin{equation}
\label{FieldEqs}
\displaystyle\frac{\ddot{p}(u)}{p(u)}+\displaystyle\frac{\ddot{q}(u)}{q(u)}=0\; ,
\end{equation}
where $\ddot{p}(u)$, $\ddot{q}(u)$ denote the second derivative of the metric functions with respect to the wave phase parameter $u$. This is a Petrov type-N metric with a Weyl scalar $\Psi_4=-\ddot{p}(u)/p(u)$ \cite{Chandrasekhar}. Alternatively, 
we can use the so-called Brinkmann's coordinates:
\begin{equation}
\label{Brinkmann}
d s^2=2 d U d V+H(U) (Y^2-Z^2) d U^2+ d Y^2+d Z^2 \; ,
\end{equation}
with $H(U)=-\Psi_4=\ddot{p}(u)/p(u)$. The coordinate transformation to arrive at Eq.\ (\ref{Brinkmann}) from Eq.\ (\ref{metric})
is $u=U$, $v=V+(\dot{p} Y^2/p+\dot{q} Z^2/q)/2$, $y=Y/p$ and $z=Z/p$. This implies that all the relevant information is
contained in the non-zero Weyl scalar, $\Psi_4$. However, to perform this coordinate transformation a couple of functions verifying Eq. (\ref{FieldEqs}), $p(u)$ and $q(u)$, must exist. 

We are interested in this paper in finding plane waves series solutions, propagating in vacuum without dispersion, according to the following criteria:

\begin{enumerate}
\item[(i)] The functions $p(u)$ and $q(u)$ are everywhere continuous and differentiable up to second order, at least. This is necessary for the definition of the
Christoffel's symbols and Ricci curvature tensor. 
\item[(ii)] The limits $\lim_{u \rightarrow \pm \infty} p(u)$ and $\lim_{u \rightarrow \pm \infty} q(u)$ exist. 
\item[(iii)] We have $p(u) \neq 0$ and $q(u) \neq 0$ for any $u \in (-\infty,\infty)$, i. e., there are no focus points.
\item[(iv)] The stress-energy pseudotensor is integrable and the energy content of the wave is finite.
\end{enumerate}

\begin{figure}
\begin{center}
\epsfig{file=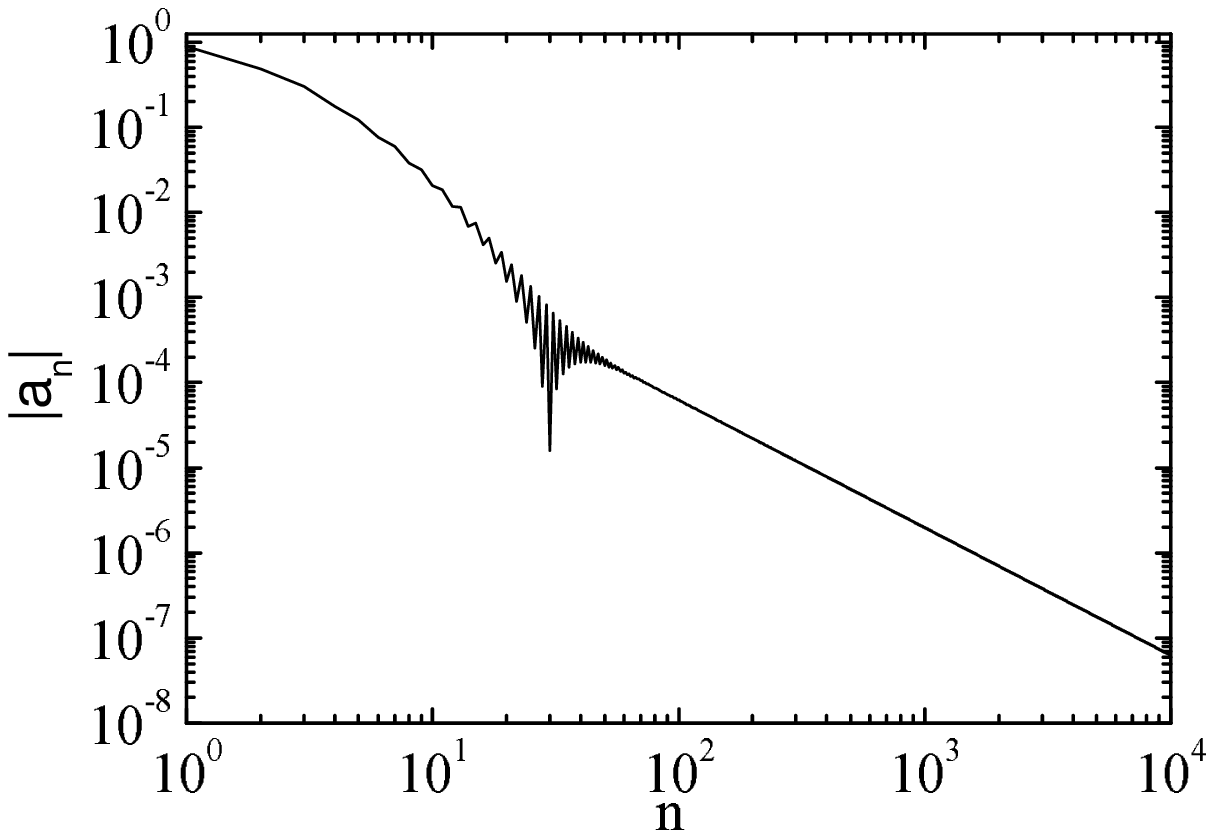,width=6cm}
\end{center}
\vspace*{8pt}
\caption{The absolute value of the first $10000$ coefficients $a_n$ for the Lorentzian gravitational wave with $\epsilon=0.9$ and $a_0=1$. Notice that a purely algebraic decay sets in for $n > 100$.\label{fig1}}
\end{figure}

These are the Bini et al. \cite{Bini} conditions particularized to this problem.
The idea of the series expansion method is to write both $p(u)$ and $q(u)$ in terms of the same function $h(u)$ which serves
as a basis for the series. We assume that $p(u)$ is a simple linear combination of $h(u)$ and a constant and that $q(u)$ is another more complicated functional:
\begin{eqnarray}
\label{pdef}
p(u)&=&1+ \epsilon\, h(u) \; , \\
\noalign{\smallskip}
\label{qdef}
q(u)&=&{\cal F}(u)=\displaystyle\sum_{n=0}^\infty\, a_n h^n(u) \; ,
\end{eqnarray}
where $\epsilon$ is a constant. From Eqs.\ (\ref{FieldEqs}), (\ref{pdef}) and (\ref{qdef}) we have:
\begin{equation}
\label{FuncEq}
\epsilon\, \ddot{h} \, {\cal F}(h)+\ddot{h}\, (1+\epsilon h)\, {\cal F}'(h)+\dot{h}^2\, (1+\epsilon h)\, {\cal F}''(h)=0\; ,
\end{equation}
where ${\cal F}'(h)$, ${\cal F}''(h)$ denote the first and second order derivative of ${\cal F}(h)$ with respect its argument, $h$. Now, for a modal series to be possible for this problem the square of the first-order derivative of the base, $\dot{h}^2(u)$, and its second-order derivative, $\ddot{h}(u)$, should be expressed as a simple polynomial in terms of the function $h(u)$ itself. A good candidate for a basis function is the Lorentzian:
\begin{equation}
\label{Lorentz}
h(u)=\displaystyle\frac{1}{1+u^2} 
\end{equation}
because it has the correct limit for $u \rightarrow \pm \infty$ and also verifies:
\begin{eqnarray}
\label{hdotL}
\dot{h}^2(u)&=&4h^3(1-h) \; , \\
\noalign{\smallskip}
\label{hddotL}
\ddot{h}(u)&=&2h^2(3-4 h) \; .
\end{eqnarray}
With these two relations in Eqs.\ (\ref{hdotL}) and (\ref{hddotL}) we can find the following non-linear differential
equation for the functional ${\cal F}(h)$:
\begin{equation}
\label{FuncEqL}
\begin{array}{rcl}
& &\epsilon (3 - 4 h){\cal F}(h)+\left(3+(3\epsilon-4)h-4\epsilon h^2\right){\cal F}'(h)\\
\noalign{\smallskip}
&+&2 h\left(1+(\epsilon-1) h- \epsilon h^2\right) {\cal F}''(h)=0\; .
\end{array}
\end{equation}
We now consider a series expansion for ${\cal F}$ as given in Eq.\ (\ref{qdef}). By substitution in Eq.\ (\ref{FuncEqL}) and 
rearranging the terms in such a way that the coefficients of the same powers of $h$ are clearly displayed we have:
\begin{equation}
\begin{array}{rcl}
& &\displaystyle\sum_{n=0}^\infty \, 3 \epsilon a_n h^n-\displaystyle\sum_{n=1}^\infty \, 4 \epsilon a_{n-1} h^n+\displaystyle\sum_{n=0}^\infty \, 3 (n+1) a_{n+1} h^n
+\sum_{n=1}^\infty\, (3 \epsilon-4) n a_n h^n\\
\noalign{\smallskip}
&-&\displaystyle\sum_{n=2}^\infty \, 4(n-1)\epsilon a_{n-1}h^n
+\displaystyle\sum_{n=1}^\infty\, 2 (n+1)n a_{n+1} h^n+\displaystyle\sum_{n=2}^\infty\, 2 (\epsilon-1) n(n-1) a_n h^n \\
\noalign{\smallskip}
&-&\displaystyle\sum_{n=3}^\infty\, 2 \epsilon (n-1)(n-2) a_{n-1} h^n=0\; .
\end{array}
\end{equation}
We can now identify the coefficients of every power of $h(u)$ with zero
in order to obtain the recurrence relations for $a_n$, $n=0,1,2,\ldots$. In particular we have:
\begin{eqnarray}
\label{a1}
a_1&=&-\epsilon a_0 \\
\noalign{\smallskip}
a_2&=&\displaystyle\frac{3}{5}\epsilon^2 a_0 \\
\noalign{\smallskip}
a_3&=&\displaystyle\frac{1}{21}\left[ 8\epsilon a_1-(13 \epsilon-12) a_2 \right]=-\displaystyle\frac{\epsilon^2}{105}\left(4+39 \epsilon\right) a_0\; ,
\end{eqnarray}
and for $n \ge 3$ the following recurrence is found:
\begin{equation}
\label{recL}
a_{n+1}=\displaystyle\frac{2\epsilon(n^2-n+2)}{(n+1)(2n+3)}a_{n-1}-
\displaystyle\frac{3\epsilon+(\epsilon-2)n+2(\epsilon-1)n^2}{(n+1)(2n+3)}a_n\; .
\end{equation}
From this relation it can be checked that the leading term in the expression of 
$a_n$ is $\epsilon^2$. So, we cannot claim that the modal series in Eq.\ (\ref{qdef}) for $u=0$ ($h(u)=1$) exhibits geometric convergence for $\epsilon < 1$. On the other hand, a numerical computation of the coefficients $a_n$ shows that they decrease with $n$. However, they decrease only algebraically with $n$ as deduced from the double-logarithmic plot in Fig.\ \ref{fig1} and, consequently, $a_{n+1}/a_n \rightarrow 1$ as $n \rightarrow \infty$. This means that convergence of the series for $q(u=0)$ in Eq.\ (\ref{qdef}) cannot be proved by the quotient test.
\section{Convergence of the Lorentzian Plane Gravitational Wave}
\label{sect_3}
We have found that the modal series expansion for the square-root of the metric of the gravitational wave, $q(u)$, is:
\begin{equation}
\label{qseries}
q(u)=\displaystyle\sum_{n=0}^\infty \, a_n h^n(u) \; .
\end{equation}
As the coefficients $\vert a_n \vert$ constitute a decreasing series for sufficiently large $n$ and $h(u) < 1$ for $u > 0$, the series in Eq.\ (\ref{qseries}) is trivially convergent for any nonzero value of $u$. We will
show in this section that $q(u=0)$, i. e., the sum of the coefficients $a_n$ is also convergent. To this end, we define a succession $b_n$ as:
\begin{equation}
\label{bRaabe}
b_n=\left(a_n/a_{n+1}-1\right)n \; .
\end{equation}
If this succession $b_n$ has a limit for $n\rightarrow \infty$ and $\lim_{n\to\infty} b_n > 1$ then, according to Raabe-Duhamel's theorem, the series for $a_n$ is convergent \cite{Arfken}. By dividing both members of the recurrence relation in Eq.\ (\ref{recL}) by $a_n$ we find that:
\begin{equation}
\label{testRaabe}
\displaystyle\frac{b_n}{n}+1=\displaystyle\frac{R(n)}{Q(n)\left(b_{n-1}/(n-1)+1\right)-P(n)}\; ,
\end{equation}
where $P(n)$, $Q(n)$ and $R(n)$ are the following second-order polynomials in $n$:
\begin{eqnarray}
P(n)&=&3\epsilon+(\epsilon-2) n+2(\epsilon-1)n^2 \\
\noalign{\smallskip}
Q(n)&=&2\epsilon(n^2-n+2)\\
\noalign{\smallskip}
R(n)&=&(n+1)(2 n+3)\; .
\end{eqnarray}
For very large values of $n$ we can assume $b_n/n \simeq b_{n-1}/(n-1)$ and Eq.\ (\ref{testRaabe}) becomes a second-order
polynomial equation for $b_n/n+1$. We can test numerically that $b_n$ is positive and the relevant root behaves as:
\begin{equation}
\label{bsol}
b_n=\displaystyle\frac{3}{2}+\displaystyle\frac{1}{(1+\epsilon) n}+{\cal O}\left(\displaystyle\frac{1}{n^2}\right)\; .
\end{equation}
Now, we have that $\epsilon \ge -1$ because, otherwise, $p(u)$ as given in Eq.\ (\ref{pdef}) becomes zero for a given $u$ and from condition (iii) we want to avoid focus points. In that case, Eq.\ (\ref{bsol}) implies that $\lim_{n\to \infty} b_n = 3/2 >1$ and the series for $q(u)$ as deduced from Eqs.\ (\ref{qdef}) and (\ref{recL}), converges. Notice that the parameter $\epsilon$ is not neccessarily small, it should only be larger than $-1$.  
\begin{figure}
\begin{center}
\epsfig{file=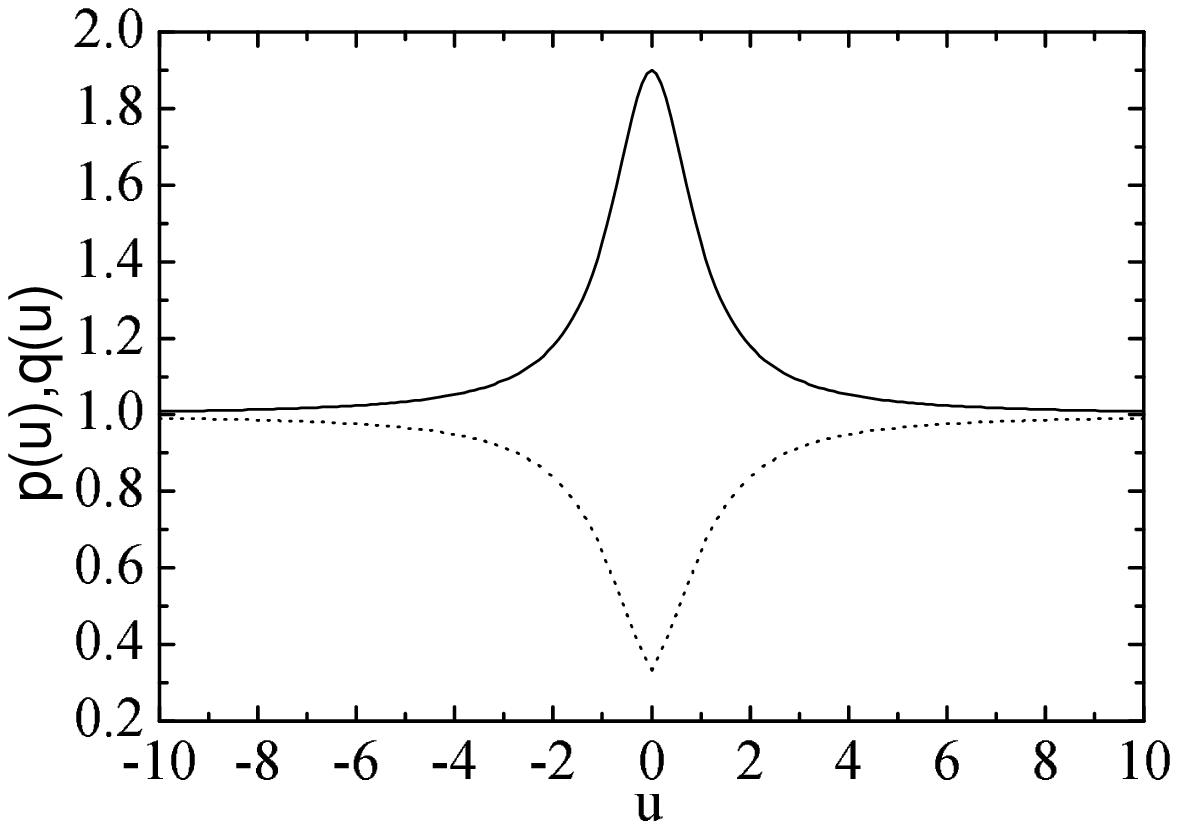,width=6cm}
\end{center}
\vspace*{8pt}
\caption{The square-root of the metric coefficients, $p(u)$ and $q(u)$ (solid and dotted lines, respectively) for the Lorentzian gravitational wave with $\epsilon=0.9$ and $a_0=1$. We have used $10^6$ terms to compute the modal series for $q(u)$ in Eq.\ (\protect\ref{qdef}). \label{fig2}}
\end{figure}
In Fig.\ \ref{fig2} we have plotted the functions $p(u)$ and $q(u)$ as given by Eqs.\ (\ref{pdef}), (\ref{qdef}) and the recurrence relation for the coefficients $a_n$ in Eq.\ (\ref{recL}) for $\epsilon=0.9$ and $a_0=1$. We notice that the passing 
gravitational wave will provoke and expansion of rods along the $y$ direction and their contraction in the $z$ direction. So, it could be
detected by laser interferometry. Nevertheless, the space-time is asymptotically Minkowskian before and after the passage of the
wave as requested in condition (ii).

It is interesting to check if a similar solution can be found for another choosing of the basis function $h(u)$. Another
convenient basis in terms of the hyperbolic cosine is:
\begin{equation}
\label{Hyperbolic}
h(u)=\displaystyle\frac{1}{1+\cosh(u)}\; ,
\end{equation}
which verifies the modal recurrences:
\begin{eqnarray}
\label{hdotH}
\dot{h}^2(u)&=&h^2 (1- 2 h) \\
\noalign{\smallskip}
\label{hddotH}
\ddot{h}(u)&=&h (1-3 h) \; ,
\end{eqnarray}
and the functional Eq.\ (\ref{FuncEq}) takes the form:
\begin{equation}
\label{FuncEqH}
\begin{array}{rcl}
& &\epsilon\, (1 - 3 h)\, {\cal F}(h)+\left(1+(\epsilon-3)h-3 \epsilon h^2\right)\,{\cal F}'(h) \\
\noalign{\smallskip}
&+& h\,\left(1+(\epsilon-2) h- 2 \epsilon h^2\right)\, {\cal F}''(h)=0\; .
\end{array}
\end{equation}
The non-linear differential equation for ${\cal F}(h)$ can be solved by the modal series expansion as before and we obtain
the coefficients:
\begin{eqnarray}
a_1&=&-\epsilon a_0 \\
\noalign{\smallskip}
a_2&=&\displaystyle\frac{3 \epsilon}{4} a_0-\displaystyle\frac{(2\epsilon-3)}{4} a_1=\displaystyle\frac{\epsilon^2}{2} a_0 \\
\noalign{\smallskip}
a_3&=&\displaystyle\frac{2 \epsilon}{3} a_1+\displaystyle\frac{5}{9} (2-\epsilon) a_2=-\displaystyle\frac{1}{18}\epsilon^2 (2+5\epsilon) \; ,
\end{eqnarray}
and for $n \ge 3$ the following general recurrence relation is found:
\begin{equation}
\label{recH}
a_{n+1}=\epsilon \displaystyle\frac{(2 n^2-3 n+4)}{(n+1)^2} a_{n-1}+\displaystyle\frac{(2-\epsilon) n^2+n-\epsilon}{(n+1)^2} a_n\; .
\end{equation}
The convergence of the series for $q(u=0)$ requires the existence of the following sum:
\begin{equation}
\label{alphadef}
\displaystyle\sum_{n=0}^\infty \, \displaystyle\frac{a_n}{2^n}=\displaystyle\sum_{n=0}^\infty \alpha_n\; .
\end{equation}
The application of Raabe-Duhamel's test in this case for the succession $b_n=(\alpha_n/\alpha_{n+1}-1)n$ proceeding in the same 
way as before yields:
\begin{equation}
\label{RaabeH}
b_n=\displaystyle\frac{3}{2}+\displaystyle\frac{2+5\epsilon}{4(2+\epsilon)} \displaystyle\frac{1}{n}+{\cal O}\left(\displaystyle\frac{1}{n^2}\right)\; .
\end{equation}
So, $\lim_{n\to\infty} b_n=3/2$ for any $\epsilon > -2$ and the series in Eq.\ (\ref{alphadef}) and also $q(u)$ as defined in Eq.\ (\ref{qdef}) with the coefficients given by the recurrence in Eq.\ (\ref{recH}) is convergent. For $\epsilon \le -2$ the metric coefficients become null for some values of $u$ and the condition (iii) is not fulfilled. So, we should restrict the solution to 
values $\epsilon > -2$.
\begin{figure}
\begin{center}
\epsfig{file=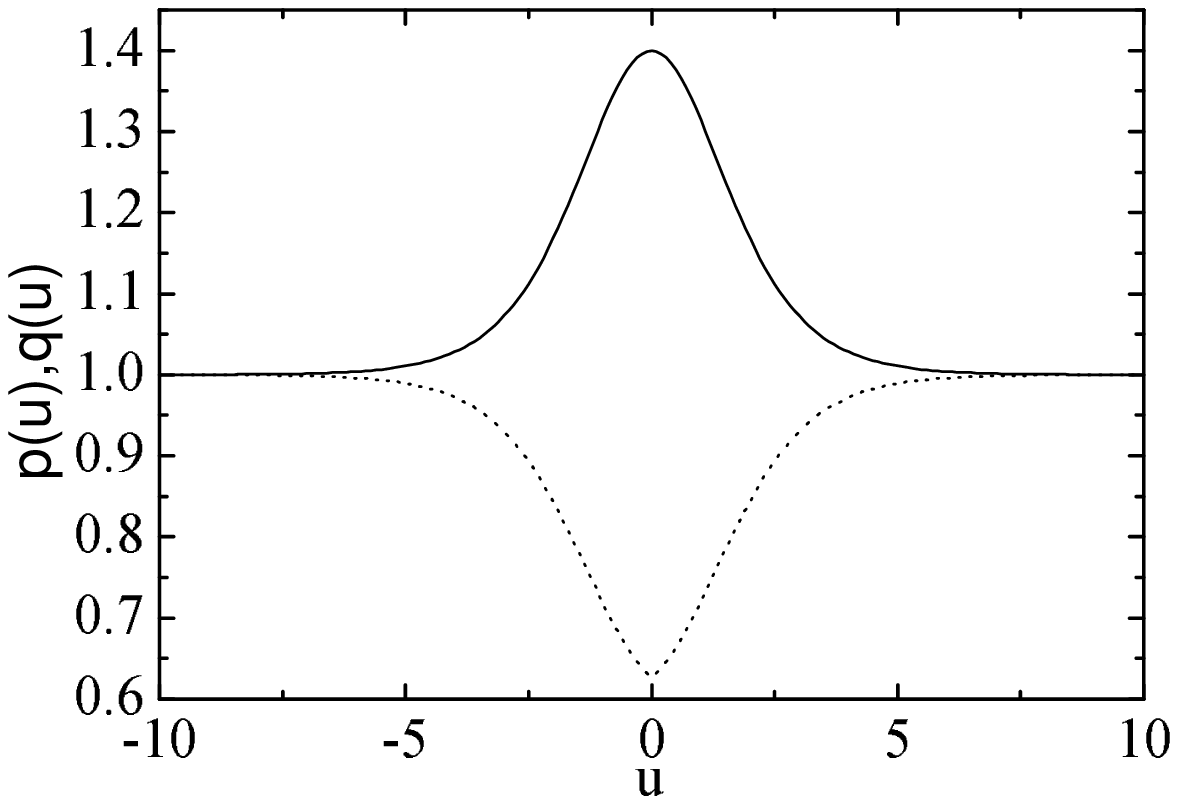,width=6cm}
\end{center}
\vspace*{8pt}
\caption{The same as Fig. \protect\ref{fig2} but for the gravitational wave defined with the basis function in Eq.\ (\protect\ref{Hyperbolic}). \label{fig3}}
\end{figure}
This gravitational wave is plotted in Fig.\ \ref{fig3} for $\epsilon=0.9$ and $a_0=1$.

Finally, we must discuss the convergence of the first and second-order derivatives of $q(u)$ because they are related
with Christoffel symbols and the Ricci tensor. The first derivative is zero for $u=0$ because the derivative of every term
in the series of Eq.\ (\ref{qdef}) is proportional to $\dot{h}(u)$. Concerning the second-order derivative the situation is
more subtle because it is given by the series:
\begin{equation}
\label{qddseries}
\ddot{q}(u)=\displaystyle\sum_{n=2}^\infty\, a_n n (n-1) h^{n-2}(u) \dot{h}^2(u)+\displaystyle\sum_{n=1}^\infty\, a_n n
h^{n-1}(u) \ddot{h}(u)\; ,
\end{equation}
For $u\neq 0$ both series in Eq.\ (\ref{qddseries}) converge by the quotient test because $\lim_{n\to\infty} a_{n+1}/a_{n}=1$ and
$h(u) < 1$. But for $u=0$ we face a problem with the second term in Eq.\ (\ref{qddseries}) which becomes proportional to $\sum_{n=1}^\infty \, n a_n$ and, consequently, diverges. However the second derivative $\ddot{q}(0)$ can still be defined as the
right and left limits of $\ddot{q}(u)$ for $u \to 0$ or, directly, from Eq.\ (\ref{FieldEqs}) as follows:
\begin{equation}
\ddot{q}(0)=-q(0)\displaystyle\frac{\ddot{p}(0)}{p(0)}\; .
\end{equation}
For example, in the case of the Lorentzian solution we obtain $\ddot{q}(u=0)=0.314565$ by computing $q(0)$ with $10^6$ terms
in the series of Eq.\ (\ref{qdef}). In comparison we have $\ddot{q}(u=0.1)=0.318772$ directly from the series in Eq.\ (\ref{qddseries}) and with the same number of terms. This proves the consistency of the solution.
\section{Energy-stress pseudotensor and the energy content of the wave}
\label{sect_4}
Gravitational waves carry a certain amount of energy. This gravitational energy can be calculated from
the so-called energy-stress pseudotensor \cite{Landau}:
\begin{equation}
\label{pseudo}
\begin{array}{rcl}
t^{ik}&=&\displaystyle\frac{c^4}{16 \pi G} \left\{ \left( 2 \Gamma_{lm}^n \Gamma_{np}^p-\Gamma_{lp}^n\Gamma_{mn}^p
-\Gamma_{ln}^n\Gamma_{mp}^p \right)\left(g^{il} g^{km}-g^{ik} g^{lm}\right) \right.\\
\noalign{\smallskip}
&+&g^{il} g^{mn} \left( \Gamma_{lp}^k \Gamma_{mn}^p+\Gamma_{mn}^k \Gamma_{lp}^p -\Gamma_{np}^k\Gamma_{lm}^p-
\Gamma_{lm}^k \Gamma_{np}^p \right)+\left[ i,k\right]\\
\noalign{\smallskip}
 &+& \left. g^{lm} g^{np} \left( \Gamma_{ln}^i \Gamma_{mp}^k-\Gamma_{lm}^i \Gamma_{np}^k\right) \right\} \; ,
\end{array}
\end{equation}
where $[i,k]$ denotes a term coinciding with the previous one save for a permutation of the indexes $i$ and $k$. $G$ is, as usual, the gravitational constant. For the Einstein-Rosen metric in Eq. (\ref{metric}) we get the following result:
\begin{equation}
\label{tpmetric}
t^{00}(u)=t^{01}(u)=-\displaystyle\frac{c^4}{8 \pi G} \left\{\dot{p}^2(u) \, q^2(u)+\dot{q}^2(u)\, p^2(u)+4 \dot{p}(u)\, p(u)\, \dot{q}(u) \, q(u)\right\} \; .
\end{equation}

\begin{figure}
\begin{center}
\epsfig{file=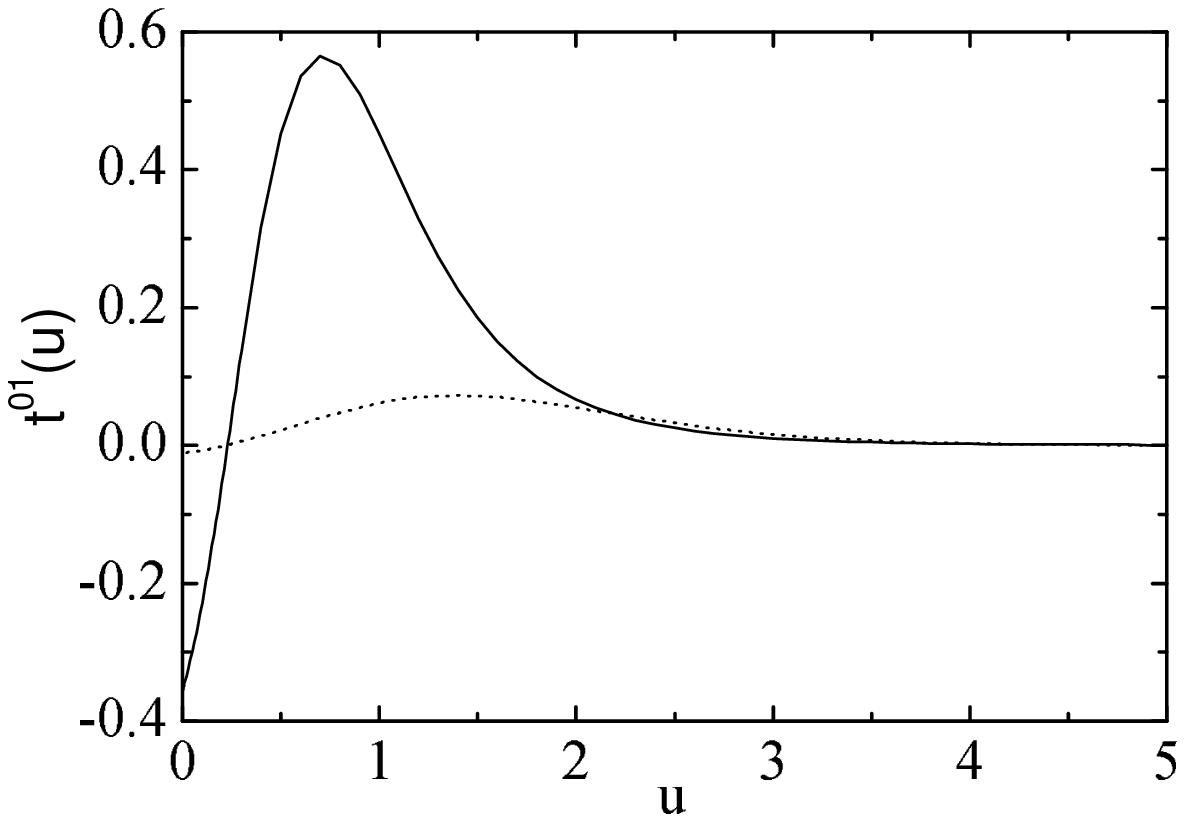,width=6cm}
\end{center}
\vspace*{8pt}
\caption{Momentum component of the stress-energy pseudotensor for the Lorentzian wave ($\epsilon=0.9$) and the Hyperbolic
wave ($\epsilon=0.8$) calculated from Eq.\ (\ref{tpmetric}) by using $10^6$ terms for the $q(u)$ series. Unit of energy density per unit length is $c^4/(16 \pi G)$ \label{fig4}}
\end{figure}

We have explicitly calculated the components of the plane wave solutions found in Sec. \ref{sect_2} and the result is plotted
in Fig. \ref{fig4}.
We notice an interesting mathematical feature of these solutions: the energy density of the plane wave may become negative
around its center. This is not physically problematic because the energy of a gravitational field is not localized and the only relevant
information is the total energy which is positive and finite (For example, $E_{\mbox{total}} = 1.06377\, c^4/(16 \pi G)$ for the Lorentzian wave). 
One of the advantages of the modal series method discussed in this paper is that the integrability of the stress-energy pseudotensor
can be easily proved. The asymptotic behaviour of the components in Eq.\ (\ref{tpmetric}) is given by:
\begin{equation}
\label{tpasymp}
t^{01}(u) \rightarrow \displaystyle\frac{\epsilon c^4}{4 \pi G} \, \dot{h}^2(u) \; , \; u \rightarrow \pm\infty
\end{equation}
where we have taken into account Eqs.\ (\ref{pdef}), (\ref{qdef}) and (\ref{a1}). In the two cases considered in Eqs.\ (\ref{Lorentz})
and (\ref{Hyperbolic}) the function $\dot{h}^2(u)$ is integrable in the domain $-\infty < u < \infty$. This is a condition we should impose to any function $h(u)$ used as seed of the modal series method.
\section{Conclusions}
\label{sect_5}   
In this paper we have proposed a modal transseries approach \cite{Costin} to the systematic analysis of plane wave solutions
of Einstein's Field Equations inspired by the success of this technique in other non-linear problems 
\cite{AcedoNARWA,AcedoPhysA,Acedo2013}. We have shown that a family of plane waves can be found in vacuum and that these
solutions satisfy requirements of physical plausibility: the space-time is Minkowskian very far away from the plane-fronted wave, the
metric coefficients are regular and everywhere different from zero, curvature is properly defined and the total gravitational energy content of the wave is finite. These waves travel without changing their shape. However, the shape of these waves is not completely determined by Einstein's Field Equations as occurs, for example, in the Korteweg-de Vries soliton as a consequence of the interplay between non-linearity and dissipation.

It would also be interesting to study the exact spherical symmetric waves and its relation to a binary pulsar or binary black hole source. Moreover,
the method developed in this paper could be applied to Non-Abelian waves in Yang-Mills theories which have been analyzed since the seventies of the past century  \cite{Coleman,Campbell,Rabinowitch}. Another interesting problem should be the analysis of collisions 
of plane waves \cite{Griffiths,Feinstein} of the type described in this paper by finding a generalization of the modal series with two fronts. 
Work along these lines will be published in future papers.

\section*{Acknowledgments}
The author gratefully acknowledges A. Feinstein for his very useful comments and a critical reading of the manuscript.

\end{document}